\documentclass[a4paper,12pt]{article}
\usepackage{amsmath,amsfonts,amssymb,color}
\usepackage{graphicx}%
\usepackage{dcolumn}%
\usepackage{bm}%
\usepackage{caption}
\usepackage{subcaption}

\begin{document}
\title{
Strain energy function for isotropic non-linear elastic incompressible solids with linear finite strain response in shear and torsion
}
\author{Robert Mangan, Michel Destrade, Giuseppe Saccomandi\\[12pt]
School of Mathematics, Statistics and Applied Mathematics,  \\ 
   NUI Galway, University Road, Galway, Ireland\\[6pt]
School of Mechanical and Materials Engineering,  \\
University College Dublin, Belfield, Dublin 4, Ireland\\[6pt]
Dipartimento di Ingegneria,\\ Universit\`{a} degli Studi di Perugia, \\ Via G. Duranti, Perugia 06125, Italy}
\date{}

\maketitle

\bigskip\bigskip

\begin{abstract}

We find the strain energy function for isotropic incompressible solids exhibiting a linear relationship between shear stress and amount of shear, and between torque and amount of twist, when subject to large simple shear or torsion deformations.
It is inclusive of the well-known neo-Hookean and the Mooney-Rivlin models, but also can accommodate other terms, as certain arbitrary functions of the principal strain invariants.
Effectively, the extra terms can be used to account for several  non-linear effects observed experimentally but not captured by the neo-Hookean and Mooney-Rivlin models, such as strain stiffening effects due to limiting chain extensibility.

\end{abstract}

\newpage


\section{Introduction}

\begin{figure}[ht!] 
        \centering
\includegraphics[width=0.5\textwidth]{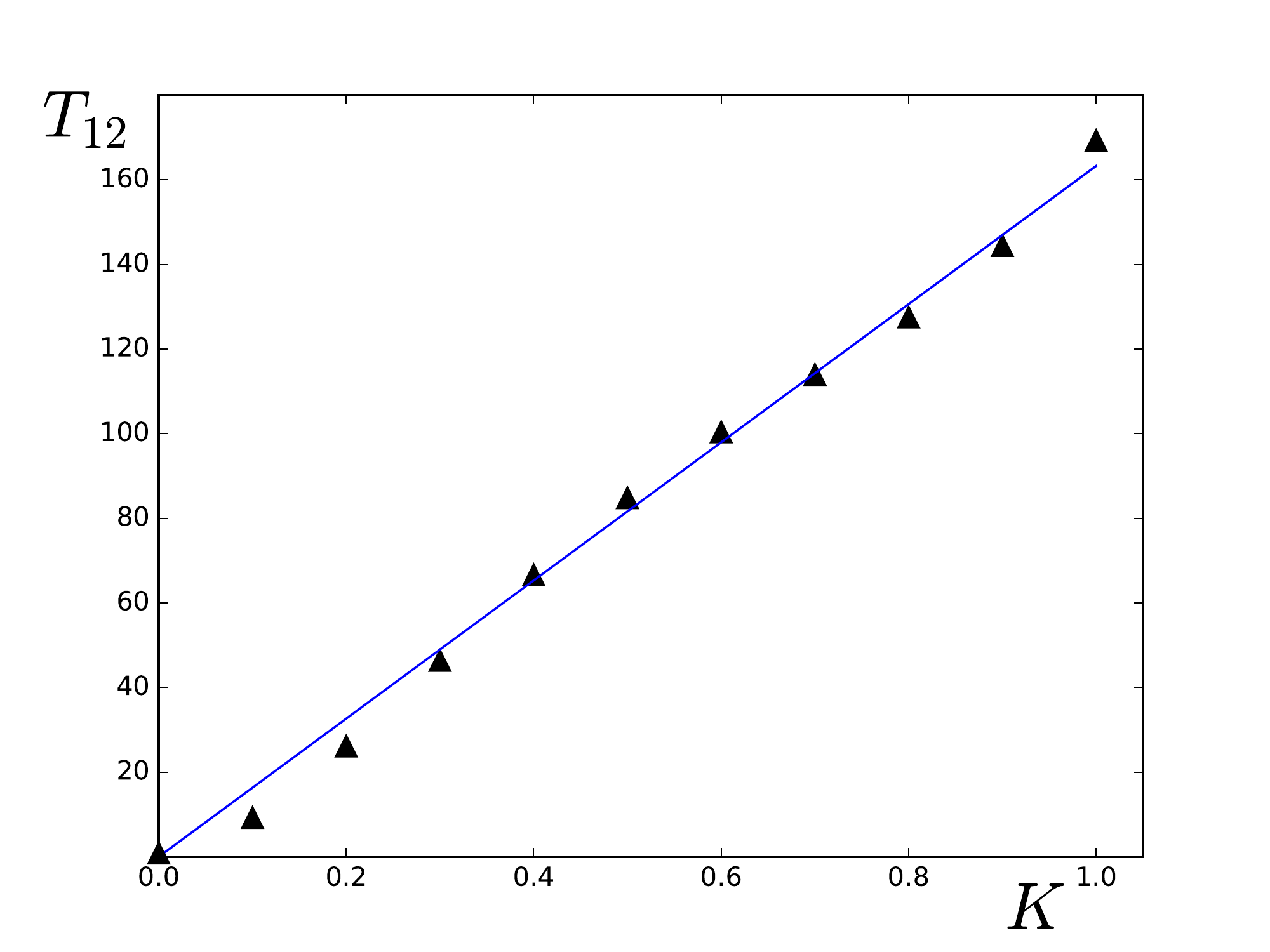}\;
\includegraphics[width=0.48\textwidth]{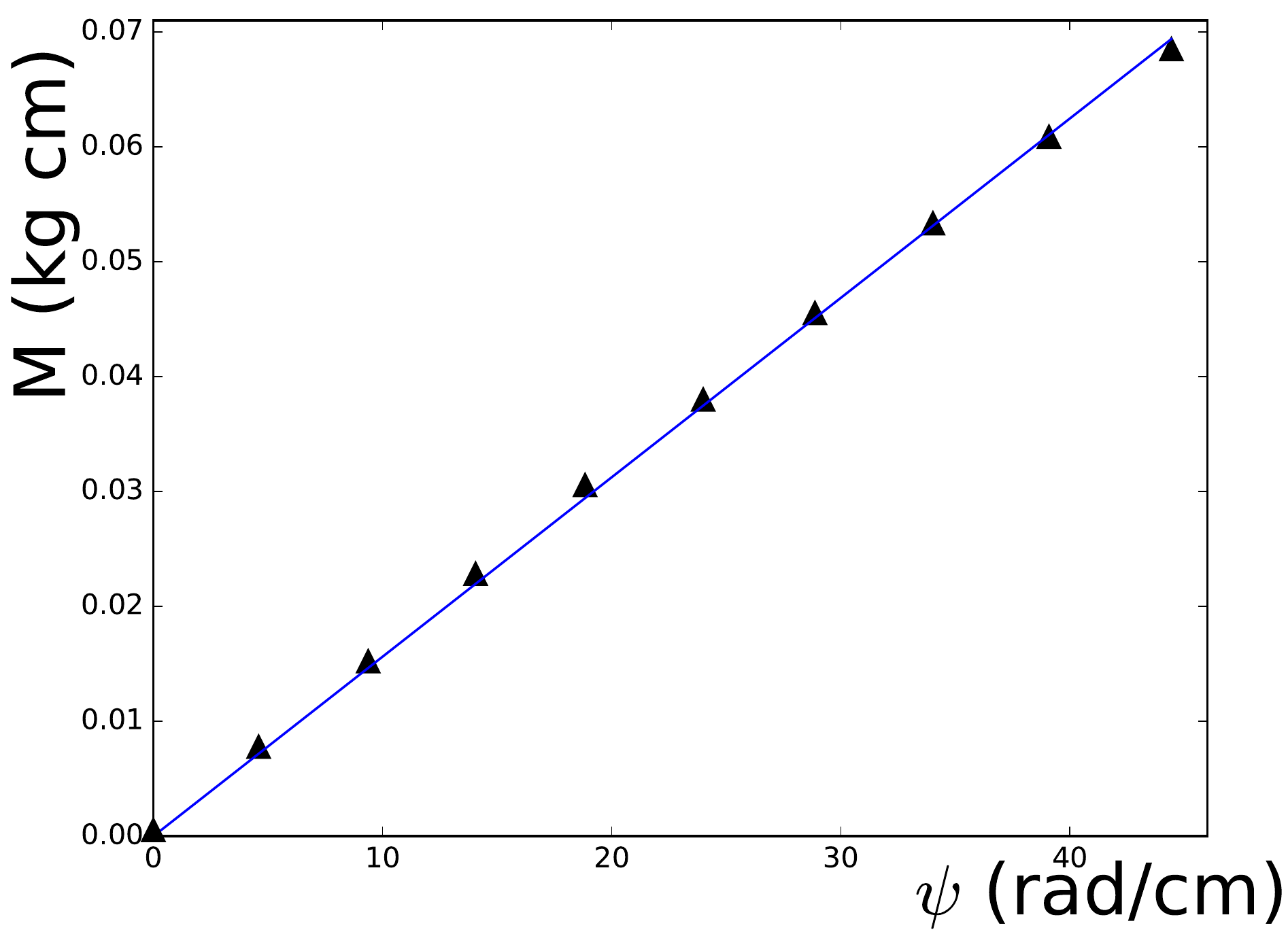}
\caption{\label{shear-plot}
(a) Shear stress response in the simple shear of porcine brain matter; experiments conducted at University College Dublin, see Ref. \cite{Dest15} for details. 
(b) Torsion of a right cylinder of rubber with radius $a= 1.27$ cm; digitized data from Ref. \cite{rivlin}.
The straight lines represent linear fittings, indicating that the shear modulus is $\mu = 163$ Pa for brain and $38.2$ kPa for rubber. 
 }
\end{figure}

Many soft incompressible materials have a linear response in shear and in torsion, including rubbers and soft tissues (FIG.\ref{shear-plot}). 
But how should that property be modeled? 
The strain energy functions that come to mind are those of the \emph{neo-Hookean} and the \emph{Mooney-Rivlin}  \cite{mooney} materials,
\begin{align} \label{nH-MR}
& W_\text{nH} = \tfrac{1}{2} C_1(I_1-3), \notag \\
& W_\text{MR}= \tfrac{1}{2}C_1(I_1-3) + \tfrac{1}{2}C_2(I_2-3),
\end{align}
respectively, where $C_1>0$, $C_2>0$ are constants, and $I_1=\text{tr}\,\bm C$, $I_2=\text{tr}(\bm C^{-1})$ are the first two principal invariants of the right Cauchy-Green deformation tensor $\bm C$. 
These models provide indeed an exact linear relationship between the Cauchy shear stress component $T_{12}$ and the amount of shear $K$, and between the torque $M$ and the twist $\psi$. 
This can be checked directly by recalling the general relationships
\begin{align}
& T_{12} = 2\left(\dfrac{\partial W}{\partial I_1} + \dfrac{\partial W}{\partial I_2}\right)K,  \notag \\[4pt] 
& M = 4 \pi \psi \int_0^a r^3 \left(\dfrac{\partial W}{\partial I_1} + \dfrac{\partial W}{\partial I_2}\right) dr,
\end{align}
(where $r$ is the radial distance and $a$ is the radius of the twisted cylinder  \cite{rivlin}) because the term in the parentheses is a constant for these two models.

However popular, these models present some significant limitations when it comes to capturing certain non-linear effects
: \emph{(1) Poynting effect:} experiments show that a normal stress develops for soft solids in simple shear \cite{Dest15}, but this cannot be captured by models that depend on $I_1$ only, like the neo-Hookean model, because their normal stress component $T_{22}= -2 (\partial W/\partial I_2)K^2$ is zero; 
\emph{(2) Strain-stiffening effect:} for large extensions, rubber-like materials stiffen rapidly and give an up-turn in the Mooney plot \cite{OgSS04}, but the neo-Hookean and the Mooney-Rivlin models only yield straight lines in that representation;
(3) \emph{Non-linear elastic response:}  with only one or  two material constants $C_1$, $C_2$ at their disposal, the neo-Hookean and Mooney-Rivlin models cannot reflect fourth-order weakly non-linear elasticity \cite{DeOg10} properly,
\begin{equation}
W_\text{4th}=\mu\,\text{tr}(\bm E^2) + \frac{A}{3}\text{tr}(\bm E^3) + D\,\text{tr}(\bm E^3),
\end{equation}
that involves three constants (here $\bm E = (\bm C -\bm I)/2$ is the Green strain tensor, $\mu$ is the infinitesimal shear modulus and $A$, $D$ are the Landau constants of third- and fourth-order elasticity)  \cite{DeOg10}.

Here we show that there exist, in fact,  more general strain energies satisfying the linearity property and able to overcome these shortcomings. 


\section{Results}


We arrive at the desired linear relationships  by enforcing that the strain energy function $W$ satisfy $\partial W/\partial I_1 + \partial W /\partial I_2 = \mathrm{constant}$. 
Furthermore, compatibility with the linear theory imposes that
$
\partial W/\partial I_1 + \partial W /\partial I_2 = \mu/2
$.
We note that the Mooney-Rivlin material \eqref{nH-MR} is a particular solution of that inhomogeneous partial differential equation, with $C_1+C_2=\mu$.
Thus the general solution may be written as
\begin{equation} 
W = W_{\text{MR}} + H(I_1,I_2),
\end{equation}
where $H$ is an arbitrary function of the two variables $I_1$, $I_2$.
Then, after substitution, we obtain a homogeneous partial differential equation for $H$,
\begin{equation} \label{pde2}
\dfrac{\partial H}{\partial I_1} + \dfrac{\partial H}{\partial I_2} = 0.
\end{equation}
The general solution of this equation is simply $H=H(I_1 - I_2)$ where $H$ remains an \emph{arbitrary function}, but now of the single variable $I_1-I_2$.
We call the corresponding class of solids, the \emph{generalized Mooney-Rivlin materials},
\begin{equation} \label{soln}
W =\tfrac{1}{2}C_1(I_1-3)+ \tfrac{1}{2}C_2(I_2-3) + H(I_1-I_2).
\end{equation}

As an illustration we consider the following example of a generalized Mooney-Rivlin material,
\begin{equation}
 \label{w-fit}
 W_\text{gMR} = W_{\text{MR}} - \tfrac{1}{2}C_3  J_m \mathrm{ln}\left(1 - \frac{I_1-I_2}{J_m}\right).
\end{equation}
This model is chosen in an attempt to capture the strain-hardening effects which occur for moderate to large extensions of rubber, and which cannot be captured by the Mooney-Rivlin model alone  \cite{OgSS04}. 
The final term of $W_\text{gMR}$ is obtained from Gent's model \cite{gent} after substituting $I_1$ by $I_1-I_2$,  and we expect that it will be able to capture limiting chain extensibility by tuning the parameter $J_m$.

For $W_\text{MR}$ and $W_\text{gMR}$ we perform curve fitting to the uni-axial extension data of Treloar  \cite{treloar}, by minimizing the relative error.  
The \emph{engineering tensile stress} $\sigma$ is given by
\begin{equation}
\sigma(\lambda) = \frac{\partial W}{\partial \lambda} = 2(\lambda - \lambda^{-2})\left(\frac{\partial W}{\partial I_1}+\lambda^{-1}\frac{\partial W}{\partial I_2} \right),
\end{equation}
where $\lambda$ is the stretch along the direction of extension, and the \emph{Mooney-plot} scales these variables as $g(z) := \sigma/(\lambda - \lambda^{-2})$ against $z:=\lambda^{-1}$.
\begin{figure}[ht!] 
        \centering
\includegraphics[width=0.48\textwidth]{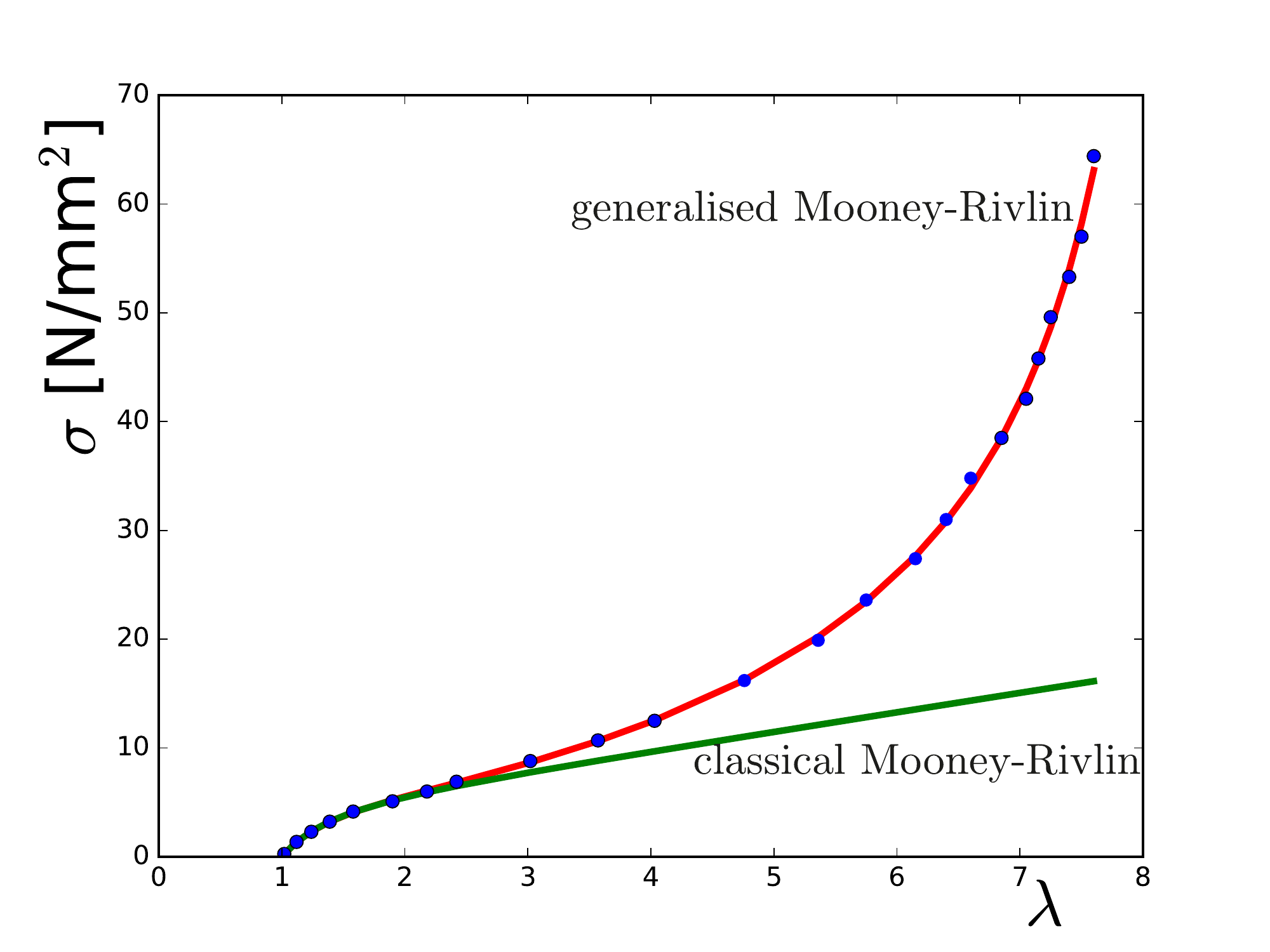}
\;
\includegraphics[width=0.48\textwidth]{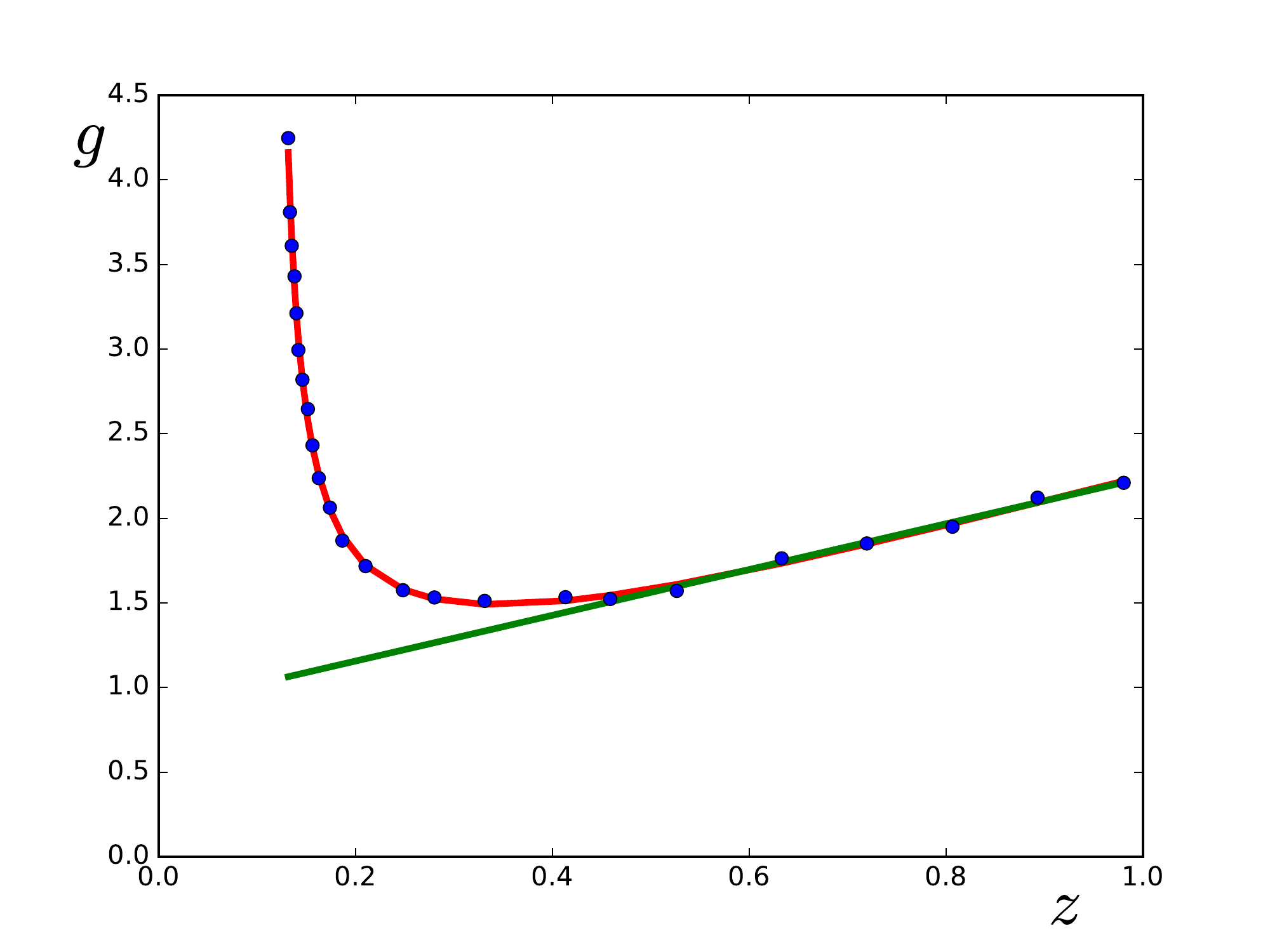}
\caption{(a) Curve-fitting to Treloar's uni-axial data \cite{treloar} (circles); Fitted $\lambda - \sigma$ curve the first seven points of the data using the Mooney-Rivlin model (lower curve) and all the data using model  $W_\text{gMR}$ (upper curve). 
(b) Mooney plots for the same models and data.}
\label{mrg-plot}
\end{figure}
For the Mooney-Rivlin material $W_\text{MR}$ the fit is made over the first seven data points only, which correspond to the linear regime in the Mooney-plot, see  \cite{OgSS04} for details and the lower (green) curves of FIG.\ref{mrg-plot}.
Over that limited range ($1 \le \lambda \lesssim 2$), it gives a maximal relative error of  1.70\% by adjusting $C_1$ and $C_2$ appropriately (explicitly, $C_1=1.7725$, $C_2=2.7042$.)
Over the entire range ($1 \le \lambda \lesssim 8$) it gives a terrible fit because it cannot accommodate the upturn in the Mooney-plot, only its early, linear part.
For the  model $W_\text{gMR}$ we perform the fitting over the entire range of stretches: we keep the same $C_1$ and $C_2$ throughout, and adjust the parameters $C_3$ and $J_m$.
The fitted curves for $W_\text{gMR}$ are plotted as the upper (red) graphs of FIG.\ref{mrg-plot}; the maximum relative error over the full range is $4.89\%$, which is well within the experimental error of Treloar. 

Further, in the fourth-order expansion \cite{DeGM10} of these models, we find the following connections for $W_\text{gMR}$,
\begin{equation}
\mu =  C_1 +C_2, \quad A=-4(C_1+2C_2 +2C_3), \quad
D= C_1 +3C_2 +4C_3.
\end{equation}


\section{Conclusion}


We note that the generalized Mooney-Rivlin models still exhibit some \emph{special} mechanical behavior. 
Indeed, when we calculate the coefficient of non-linearity of non-linear acoustics \cite{Zabo04} $\beta = (\mu+A/2+D)/(2\mu)$ we obtain $\beta=0$, not only for the specific example \eqref{w-fit} but for  the entire class of generalized Mooney-Rivlin materials. 
As a result, these materials cannot be used to model non-linear shear wave propagation.  
Moreover,  because $\beta=0$, they will not predict unbounded growth for the bending moment of a rectangular block with increasing values of the product of the block aspect ratio by the bending angle \cite{KaHo08,DeGM10}. 
To overcome these problems associated with the linearity of the models in shear and in torsion, we have to recognize that the linearity property exists only over a limited range of stretches, and we then have to undertake a completely different approach to the modeling, as explained in a recent contribution on mathematical models of rubber-like materials \cite{DSS}. 

Nonetheless, the class of generalized Mooney-Rivlin materials achieves Mooney's \emph{aspiration} \cite{mooney} of a model obeying Hooke's law in shear over \emph{a wide range of deformation} and for which \emph{neither the force-elongation nor the stress-elongation relationship agrees with Hooke's law}  in simple extension.  The models proposed improve on the predictions of the Mooney-Rivlin model  in simple extension over the whole range of admissible deformations and hence provide a rich alternative to the model first proposed by Mooney \cite{mooney} and later re-elaborated by Rivlin and co-authors \cite{rivlin}. 


\section*{Acknowledgements}


\noindent
We are thankful to Jerry Murphy for helpful discussions and to Badar Rashid for the simple shear experiment of FIG.\ref{shear-plot}. 
RM gratefully acknowledges the funding of his PhD by a scholarship from the Irish Research Council. 
The research of GS is partially funded by GNFM of Istituto Nazionale di Alta Matematica.


\end{document}